\documentclass[showpacs,aps,prd,nofootinbib,floatfix,amsmath,amssymb]{revtex4}
\usepackage{graphicx}

\begin{document}


\title{Effective Lagrangian approach to Higgs-mediated FCNC top quark decays}
\author{Adriana Cordero--Cid}
\email[E-mail:]{lcordero@venus.ifuap.buap.mx}
\affiliation{Instituto de F\'{\i}sica, Benem\'erita Universidad
Aut\'onoma de Puebla, Apartado Postal J-48, 72570 Puebla, Pue.
M\'exico}
\author{M. A. P\' erez}
\email[E-mail:]{mperez@cinvestav.fis.mx} \affiliation{Departamento
de F\'\i sica, CINVESTAV, Apartado Postal 14-740, 07000, M\'exico,
D. F., M\'exico}
\author{G. Tavares--Velasco}
\email[E-mail:]{gtv@fcfm.buap.mx}
\author{J. J. Toscano}
\email[E-mail:]{jtoscano@fcfm.buap.mx} \affiliation{Facultad de
Ciencias F\'\i sico Matem\'aticas, Benem\'erita Universidad
Aut\'onoma de Puebla, Apartado Postal 1152, Puebla, Pue., M\'
exico}

\date{\today}

\begin{abstract}
The flavor changing neutral current (FCNC) transitions $t\to q'H$
and $t\to q'V_i$ $(V_i=\gamma, g, Z)$ are studied in the context
of the effective Lagrangian approach. We focus on the scenario in
which these decays are predominantly induced by new physics
effects arising from the Yukawa sector extended with dimension-six
$SU_L(2)\times U_Y(1)$-invariant operators, which generate the
most general CP-even and CP-odd $tq'H$ vertex at the tree level.
For the unknown coefficients, we assume a slightly modified
version of the Cheng-Sher ansatz. We found that the branching
ratio for the Higgs-mediated FCNC $t\to q'V_i$ decays are enhanced
by two or three orders of magnitude with respect to the results
expected in models with extended Higgs sectors, such as the
general two-Higgs doublet model. We discuss the possibilities of
detecting this class of decays at the LHC.

\end{abstract}

\pacs{}

\maketitle

\section{Introduction}
\label{int}The fact that the top quark is the only known fermion
whose mass is comparable to the electroweak symmetry breaking
scale suggests that it may be more sensitive to new physics
effects than the remaining known fermions. Furthermore, the new
dynamic effects are likely to be more evident in those top quark
processes that are forbidden or strongly suppressed in the
standard model (SM). Therefore, the study of the flavor changing
neutral current (FCNC) transitions of the top quark could be the
clue to detect new physics effects \cite{Review}. In the SM, the
$t\to c\gamma$, $t\to cZ$, and $t \to cH$ decays have branching
ratios of the order of $10^{-13}$ \cite{Perez,MP}, which are too
small to be detected at collider experiments. Any signal of such
transitions will thus represent a neat evidence of new physics. So
far, most studies on FCNC top quark transitions have focused on
the $t\to cV_i$ decays, with $V_i=\gamma, Z, g$. However, since
the Higgs boson $H$ may be as heavy as the top quark, the
electroweak symmetry would also be maximally broken by this
particle, thereby opening the possibility that some Higgs-mediated
FCNC effects could be observed in future experiments. The study of
this scenario is encouraged by the possibility of a common source
responsible for both symmetry breaking and flavor changing
effects, along with the experimental prospects: a copious
production of top quark events is expected at the CERN large
hadron collider (LHC).

Although scalar-mediated FCNC effects are strongly suppressed in
the SM, they may be more significant in some of its extensions.
For instance, the general two-Higgs doublet model (THDM-III) has
the simplest extended Higgs sector that naturally introduces
scalar-mediated FCNC effects at the tree level
\cite{ORIGINAL,SAVAGE}. As a result, in that model the
scalar-mediated top quark transitions may have branching ratios
several orders of magnitude larger that those predicted by the SM
\cite{REINA1,REINA2,THDMH,DPTT}. A similar conclusion was reached
in the case of Higgs-mediated lepton-flavor violating processes
\cite{THDML,DT}. Large Higgs-mediated FCNC effects can also be
naturally induced in other SM extensions, such as the minimal
supersymmetric standard model \cite{SUSYH}.

As already mentioned, FCNC effects can be induced in theories with
an extended scalar sector or a larger gauge group
\cite{ORIGINAL,THDMH,THDML}. In the present work we are interested
in those new physics effects that induce FCNC top quark
transitions predominantly mediated by the SM Higgs boson. We will
take a similar approach to that used in Ref. \cite{DT} to study
Higgs boson-mediated lepton flavor violating effects. We found it
convenient to perform a model independent study by means of the
effective Lagrangian approach (ELA), which is appropriate to
investigate new physics effects in processes that are forbidden or
strongly suppressed in the SM \cite{Buchmuller,EW}. In this scheme
no new degrees of freedom are introduced but the SM ones. We will
assume an effective Lagrangian composed of only one Higgs
doublet,\footnote{We may also consider an effective Lagrangian
composed of an extended scalar sector \cite{Wudka}, but the
corresponding FCNC effects would have a rather different origin to
the one of those described in this work.} which will be taken as
the one responsible for FCNCs, which in turn may arise from
virtual effects of heavy particles lying beyond the Fermi scale.
Moreover, motivated by the role played by the Yukawa sector in
flavor physics, we will assume that the main source of
Higgs-mediated FCNC top quark transitions is the Yukawa sector
extended with dimension-six operators. We would like to emphasize
that the scenario that we are interested in is different to that
arising in models with extended scalar sectors
\cite{SAVAGE,REINA2}, in which the scalar FCNCs are induced at the
tree level by the presence of additional Higgs multiplets rather
than by virtual heavy particles.

In the scenario already described, the $tq'V_i$ vertices
($q'=u,c$) would be necessarily induced at the one-loop level via
the anomalous $tq'H$ coupling. In fact, all of the  $tq'V_i$
couplings but the $tq'Z$ one can only arise at the one-loop level
in any renormalizable theory. They can be conveniently
parametrized through the following effective Lagrangian
\cite{HanVertices}
\begin{eqnarray}
\label{el} \mathcal{L}&=&\bar{t}\Bigg\{\frac{ie}{2m_t}
\,(\kappa_{tq'\gamma}+i\widetilde{\kappa}_{tq'\gamma}\gamma_5)\sigma_{\mu
\nu}\,F^{\mu
\nu}+\frac{ig_s}{2m_t}\,(\kappa_{tq'g}+i\widetilde{\kappa}_{tq'g})\sigma_{\mu
\nu}\frac{\lambda^a}{2}\,G^{\mu \nu}_a +\nonumber
\\
&&\frac{g}{2c_W}\Big[\gamma_\mu(v_{tq'Z}+a_{tq'Z}\gamma_5)Z^\mu +
\frac{i}{2m_t}(\kappa_{tq'Z}+
i\widetilde{\kappa}_{tq'Z}\gamma_5)\sigma_{\mu \nu}Z^{\mu
\nu}\Big]\Bigg\}q'+{\rm H.c.}
\end{eqnarray}
The main goal of this work is to assess the impact of the $tq'H$
vertex on the $tq'V_i$ vertices and thus on the rare $t\to q'V_i$
decays. Since the effective Yukawa sector can generate the most
general $tq'H$ coupling, it would induce both the CP-even and
CP-odd $tq'V_i$ couplings at the one-loop level. We believe that
this is an interesting scenario as it is expected that the
Higgs-top dynamics is sensitive to new physics effects. In
particular, FCNC effects would be favored by the involved mass
scales: we will show below that even though the $t\to q'V_i$
decays are induced at the one-loop level by the $tq'H$ vertex, the
corresponding amplitudes are unsuppressed because both the
external and the internal scales, $m_t$ and $m_H$, are expected to
be of the same order of magnitude. The most general $tq'H$ vertex
involves two unknown parameters: a CP-even one, $\epsilon_{tq'H}$,
and a CP-odd one, $\widetilde{\epsilon}_{tq´H}$, whose order of
magnitude can be constrained from low-energy data. Since the
bounds on these parameters turn out to be somewhat loose, we will
adopt a slightly modified version of the Cheng-Sher ansatz to
estimate them and predict the rates of the FCNC top quark decays.

The rest of this paper has been organized as follows. In Sec.
\ref{model} we will derive the $tq'H$ vertex from the most general
effective Yukawa-type operators of dimension six, which
simultaneously incorporate both FCNC and CP-violating effects.
Section \ref{loop} is devoted to the calculation of the $tq'H$
contribution to the $tq'V_i$ couplings. In Sec. \ref{Br}, we
evaluate the rates of the FCNC top quark decays and discuss the
results. Particular attention will be paid to emphasize the
differences between the scenario discussed in this work and some
specific models. Finally, the conclusions are presented in Sec.
\ref{c}.

\section{Effective Yukawa sector and Higg-mediated FCNC effects}
\label{model}It is well-known that the SM Yukawa sector is both
flavor- and CP-conserving. FCNC effects can arise at tree level in
any renormalizable Yukawa sector if more scalar fields are
incorporated. However, it is not necessary to introduce new
degrees of freedom to generate both FCNC and CP-violating effects
if Dyson power counting criterion of renormalizability is not
granted as a fundamental principle when constructing the
Lagrangian. Indeed, it is only necessary to extend the Yukawa
sector with dimension-six operators to induce the most general
couplings of the Higgs boson to the quarks. A Yukawa sector with
those features, which respects the $SU_L(2)\times U_Y(1)$
symmetry, has the following structure \cite{Shalom}

\begin{equation}
\mathcal {L}^Y_{eff}=-Y^d_{ij}(\bar{Q}_i\Phi
d_j)-Y^u_{ij}(\bar{Q}_i\widetilde{\Phi}u_j)
-\frac{\alpha^d_{ij}}{\Lambda^2}(\Phi^\dag \Phi)(\bar{Q}_i\Phi
d_j)-\frac{\alpha^u_{ij}}{\Lambda^2}(\Phi^\dag
\Phi)(\bar{Q}_i\widetilde{\Phi}u_j)+{\rm H.c.},
\end{equation}
where $Y_{ij}$, $Q_i$, $\Phi$, $d_i$, and $u_i$ stand for the
usual Yukawa constants, the left-handed quark doublet, the Higgs
doublet, and the right-handed quark singlets of down and up type,
respectively. The $\alpha_{ij}$ constants, which parametrize the
details of the underlying physics, could be determined once the
fundamental theory is known. Since the dimension-six operators can
be generated at the tree level in the fundamental theory
\cite{EW}, they would not be suppressed by the loop factors as it
occurs with those operators that generate the gluon- and
photon-mediated FCNC top quark transitions at the tree level
\cite{HanVertices,FHTT}.

After spontaneous symmetry breaking, $\mathcal {L}^Y_{eff}$ can be
diagonalized as usual via the unitary matrices $V^{u,d}_L$ and
$V^{u,d}_R$, which relate gauge states to mass eigenstates. The
diagonalized Lagrangian can be written in the unitary gauge as
follows:
\begin{eqnarray}
\label{Lagrangian}
\mathcal{L}^Y_{eff}&=&-\left(1+\frac{gH}{2m_W}\right)\left(\bar{D}M^dD+\bar{U}M^uU\right)
-H\left[1+\frac{gH}{4m_W}\left(3+\frac{gH}{2m_W}\right)\right]\nonumber
\\
&&\times
\Big[\bar{D}(\Omega^dP_R+\Omega^{d\dag}P_L)D+\bar{U}(\Omega^uP_R+\Omega^{u\dag}P_L)U\Big],
\end{eqnarray}
where $M^{d,u}$ are diagonal mass matrices, whereas $U^T=(u,c,t)$
and $D^T=(d,s,b)$ are vectors in flavor space. The first term in
this expression corresponds to the usual Yukawa sector of the SM.
The $\Omega^{u,d}$ matrices, which represent the new physics
effects, are given by
\begin{equation}
\Omega^{u,d}=V^{u,d}_L\frac{1}{\sqrt{2}}\left(\frac{v}{\Lambda}\right)^2\alpha^{u,d}
V^{\dag u,d}_R.
\end{equation}
To generate scalar-mediated FCNC effects at the tree level it is
assumed that neither $Y^{u,d}$ nor $\alpha^{u,d}$ are diagonalized
by the $V^{u,d}_L$ and $V^{u,d}_R$ matrices, which should only
diagonalize the sum $Y^{u,d}+\alpha^{u,d}$. Under this assumption,
mass and interaction terms would not be simultaneously
diagonalized as it occurs in the renormalizable sector. In
addition, if ${\Omega^{u,d}}^\dag \neq \Omega^{u,d}$ the
Lagrangian (\ref{Lagrangian}) would induce both the CP-even and
CP-odd couplings of the Higgs boson to quark pairs. The
corresponding Lagrangian for the up sector can be written as

\begin{equation}
\mathcal{L}_{U_iU_jH}=-H\bar{U_i}(\epsilon_{U_i
U_jH}+i\widetilde{\epsilon}_{U_iU_jH}\gamma_5)U_j+{\rm H.c.},
\end{equation}
where $\epsilon_{U_iU_jH}=Re(\Omega^u_{ij})$ and
$\widetilde{\epsilon}_{U_iU_jH}=Im(\Omega^u_{ij})$. In the
following, only the first order terms in $\epsilon_{U_iU_jH}$ and
$\widetilde{\epsilon}_{U_iU_jH}$ will be retained in the
transition amplitudes.

To estimate the FCNC top quark decays we need to assume some
values for $\epsilon_{tq'H}$ and $\widetilde{\epsilon}_{tq'H}$. It
turns out that the bounds that can be obtained from experimental
data are somewhat loose. For instance, the $\epsilon_{tq'H}$
parameter can be bounded using the experimental data on the
$\kappa_{tq'\gamma}$ and $v_{tq'Z}$ parameters \cite{Abe}. The
explicit calculation leads to the bound $|\epsilon_{tq'H}|<O(10)$,
which would yield overestimated predictions for the rates of the
FCNC top quark decays. We will adopt instead the Cheng-Sher
ansatz, which will be slightly modified by introducing the new
physics scale $\Lambda$ instead of the Fermi one $v$. We will
assume that
\begin{equation}
\epsilon_{tq'H}=\lambda_{tq'}\frac{\sqrt{m_tm_{q'}}}{\Lambda},
\end{equation}

\begin{equation}
\widetilde{\epsilon}_{tq'H}=\widetilde{\lambda}_{tq'}\frac{\sqrt{m_tm_{q'}}}{\Lambda}.
\end{equation}
In this way, a hierarchy given by the relevant scales is
automatically introduced, which incorporates the two quark masses
and the new physics scale. As usual, we will take
$\lambda_{tq'}\sim \widetilde{\lambda}_{tq'} \sim 1$. We find it
natural to introduce the new physics scale $\Lambda$ in order to
consider the most general scenario, including those cases that
could arise beyond renormalizable theories. After using this
ansatz for $\epsilon_{tq'H}$ and $\widetilde{\epsilon}_{tq'H}$,
the only free parameters are the SM Higgs boson mass and the
$\Lambda$ scale. Although effective field theories require
$\Lambda$ to be larger than $v$, it needs not to be much larger.
This means that the inclusion of higher dimension operators is
necessary if a high precision is to be achieved. The scenario in
which $\Lambda$ is close to $v$ is interesting and will be
explored when analyzing the FCNC top quark transitions.

\section{$tq'H$  contribution to the $tq'V_i$ couplings}
\label{loop}

\begin{figure}[!htb]
\centering
\includegraphics[width=3in]{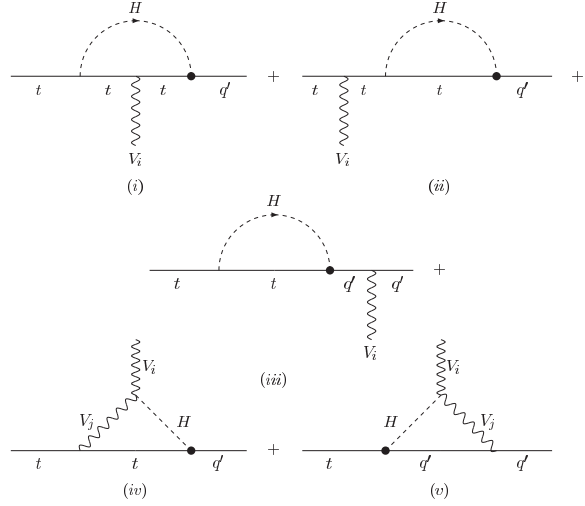}
\caption{\label{Diagrams} Feynman diagrams contributing to the
on-shell $tq'V_i$ $(V_i=\gamma, g, Z$ and $q'=u,c)$ vertices in
the $m_{q'}=0$ approximation. The diagrams (iv) and (v), which
only contribute to the $tq'Z$ vertex, were evaluated using the
unitary gauge. }
 \end{figure}

In the $m_{q'}=0$ approximation, which will be used whenever
possible, the $tq'V$ vertex arises from the Feynman diagrams shown
in Fig. \ref{Diagrams}. The $tq'\gamma$ and $tq'g$ couplings are
induced by the (i)-(iii) diagrams, whereas the $tq'Z$ vertex
receives additional contributions from the (iv) and (v) graphs.
The contributions to the $\kappa_{tq'\gamma}$ and
$\widetilde{\kappa}_{tq'\gamma}$ coefficients can be written as
follows:
\begin{equation}
\kappa_{tq'\gamma}=-\frac{2Q_t\sqrt{\alpha}}{(4\pi)^{3/2}s_{2W}x}\,\epsilon_{tq'H}\,A(y)
\end{equation}
\begin{equation}
\widetilde{\kappa}_{tq'\gamma}=-\frac{2iQ_t\sqrt{\alpha}}{(4\pi)^{3/2}s_{2W}x}\,\widetilde{\epsilon}_{tq'H}\,A(y),
\end{equation}
with $Q_t=2/3$, $x=m_Z/m_t$, $y=m_H/m_t$, and $A(y)$ the loop
function given by
\begin{equation}
A(y)=\frac{1}{2}+2m^2_tC_0+(2-y^2)\left(B_0(2)-B_0(1)\right).
\end{equation}
where $C_0=C_0(m^2_t,0,0,m^2_ty^2,m^2_t,m^2_t)$,
$B_0(1)=B_0(0,m^2_t,m^2_ty^2)$, and
$B_0(2)=B_0(m^2_t,m^2_t,m^2_ty^2)$ are Passarino-Veltman scalar
functions written in the usual notation.

The contributions to $\kappa_{tq'g}$ and
$\widetilde{\kappa}_{tq'g}$, follow easily after the
$g_s\lambda^\alpha/2$ factor is included into the
$\kappa_{tq'\gamma}$ and $\widetilde{\kappa}_{tq'\gamma}$
coefficients. As for the coefficients associated with the $tq'Z$
vertex, they are

\begin{equation}
v_{tq'Z}=-\frac{\sqrt{\alpha}}{(4\pi)^{3/2}x
s_{2W}}\left[g^{q'}_V\epsilon_{tq'H}A_1(x,y)-ig^{q'}_A\widetilde{\epsilon}_{tq'H}A_2(x,y)\right
],
\end{equation}
\begin{equation}
a_{tq'Z}=-\frac{\sqrt{\alpha}}{(4\pi)^{3/2}x
s_{2W}}\left[g^{q'}_A\epsilon_{tq'H}A_2(x,y)-ig^{q'}_V\widetilde{\epsilon}_{tq'H}A_1(x,y)\right]
\end{equation}
\begin{equation}
\kappa_{tq'Z}=\frac{\sqrt{\alpha}}{(4\pi)^{3/2}x s_{2W}}\left
[g^{q'}_V\epsilon_{tq'H}A_3(x,y)+ig^{q'}_A\widetilde{\epsilon}_{tq'H}A_4(x,y)\right],
\end{equation}
\begin{equation}
\widetilde{\kappa}_{tq'Z}=\frac{i\sqrt{\alpha}}{(4\pi)^{3/2}x
s_{2W}}\Big [g^{q'}_A\epsilon_{tq'H}A_4(x,y)
+ig^{q'}_V\widetilde{\epsilon}_{tq'H}A_3(x,y)\Big ],
\end{equation}
where $g^{q'}_V=1/2-(4/3)s^2_W$ and $g^{q'}_A=1/2$. The $A_i$
functions are given by

\begin{equation}
A_1(x,y)=\frac{x^2}{\chi ^3}\left[f^1_0+\sum^8_{i=1,i\neq
3,6}f^1_iB_0(i)+m^2_t\sum^3_{i=1}g^1_iC_0(i)\right],
\end{equation}
\begin{equation}
A_2(x,y)=\frac{1}{\chi
^3}\left[f^2_0+\sum^8_{i=1}f^2_iB_0(i)+m^2_t\sum^3_{i=1}g^2_iC_0(i)\right],
\end{equation}
\begin{equation}
A_3(x,y)=\frac{1}{\chi ^3}\left[f^3_0+\sum^8_{i=1,i\neq
3}f^3_iB_0(i)+m^2_t\sum^3_{i=1}g^3_iC_0(i)\right],
\end{equation}
\begin{equation}
A_4(x,y)=\frac{1}{\chi
^3}\left[f^4_0+\sum^8_{i=1}f^4_iB_0(i)+m^2_t\sum^3_{i=1}g^4_iC_0(i)\right],
\end{equation}
with $\chi =1-x$. The $B_0(i)$ and $C_0(i)$ scalar functions
together with the $f^a_i$ and $g^a_i$ coefficients are presented
in an Appendix. As a crosscheck, we have verified that the
amplitude for the $t\to q'Z$ decay reproduces that associated with
$t\to q'\gamma$ when $g^{q'}_V=1$, $g^{q'}_A=0$, and $x=0$.

It is straightforward to show that $\sum^8_{i=1}f^j_i=0$ for
$j=1\ldots 4$, which means that all of the amplitudes are free of
ultraviolet divergences. Thus, after introducing the Yukawa-like
operators it is not necessary to use a renormalization scheme.
This is due to the fact that the $tq'H$ vertex has a
renormalizable structure.

\begin{figure}[!htb]
\centering
\includegraphics[width=3in]{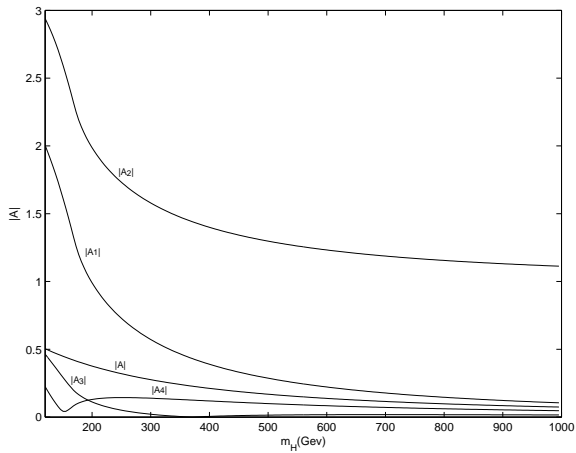}
\caption{\label{Amplitudes_1} The behavior of the $t\to cV_i$ loop
amplitudes for a very heavy Higgs boson.}
\end{figure}

\begin{figure}[!htb]
\centering
\includegraphics[width=3in]{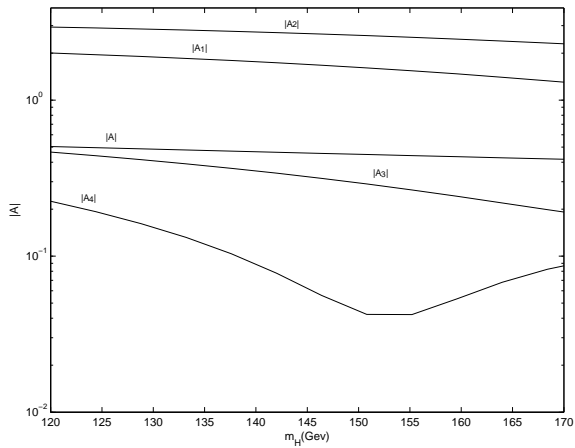}
\caption{\label{Amplitudes_2} The same as in Fig.
\ref{Amplitudes_1}, but for 120 GeV $\le m_H\le $170 GeV. }
\end{figure}

It is interesting to analyze the dependence of the loop amplitudes
on the Higgs boson mass $m_H$. They are shown in Fig.
\ref{Amplitudes_1} for very large $m_H$ and in Fig.
\ref{Amplitudes_2} in the intermediate $m_H$ regime. The
decoupling nature of the loop amplitudes is evident in Fig.
\ref{Amplitudes_1}. Also, since these amplitudes vary smoothly
with increasing $m_H$, as observed in Fig. \ref{Amplitudes_2}, the
corresponding decay widths will show the same behavior. We can
also infer the sensitivity of the $tq'V_i$ vertices to the $tq'H$
coupling. From Fig. \ref{Amplitudes_2} we can observe that
$|A_1|\sim 4|A|$, $|A_2|\sim 6|A|$, $|A_3|\sim |A|$, and
$|A_4|\sim |A|/2$, with $|A|$ ranging from 0.5 to 0.42. This means
that the coefficients $a_{tq'Z}$ and $v_{tq'Z}$ associated with
the $tq'Z$ vertex are the most sensitive to the $tq'H$ vertex.

\section{Numerical results and discussion}\label{Br}

We turn to the numerical results for the $t\to q'V_i$ and $t\to
q'H$ branching ratios. In terms of the coefficients of the
effective Lagrangian (\ref{el}), the branching ratios can be
written as

\begin{equation}
Br(t\to q'\gamma)=\frac{\alpha}{2} \left(\frac{
m_t}{\Gamma_t}\right)\left(|\kappa_{tq'\gamma}|^2+|\widetilde{\kappa}_{tq'\gamma}|^2\right),
\end{equation}
\begin{equation}
Br(t\to
q'g)=\frac{2\alpha_s}{3}\left(\frac{m_t}{\Gamma_t}\right)\left(|\kappa_{tq'g}|^2+|\widetilde{\kappa}_{tq'g}|^2\right),
\end{equation}
where $\Gamma_t$ is the total top quark width.

As for the $t\to q'Z$ decay, its branching ratio can be written as

\begin{eqnarray}
Br(t\to q'Z)&=&\frac{\alpha}{4
s^2_{2W}}\left(\frac{m_t}{\Gamma_t}\right)(1-x^2)^2\Bigg\{\frac{1+2x^2}{x^2}\left(|v_{tq'Z}|^2+|a_{tq'Z}|^2\right)\nonumber
\\
&&-6Re(v_{tq'Z}\kappa^*_{tq'Z}+a_{tq'Z}\widetilde{\kappa}^*_{tq'Z})
+(2+x^2)\left(|\kappa_{tq'Z}|^2+|\widetilde{\kappa}_{tq'Z}|^2\right)\Bigg\}.
\end{eqnarray}
whereas for the $t\to q'H$ decay we have

\begin{equation}
Br(t\to
q'H)=\frac{1}{16\pi}\left(\frac{m_t}{\Gamma_t}\right)\left(1-y^2\right)^2
\left(|\epsilon_{tq'H}|^2+|\widetilde{\epsilon}_{tq'H}|^2\right),
\end{equation}
which is a tree level prediction in the effective theory.

\begin{figure}[!htb]
\centering
\includegraphics[width=3in]{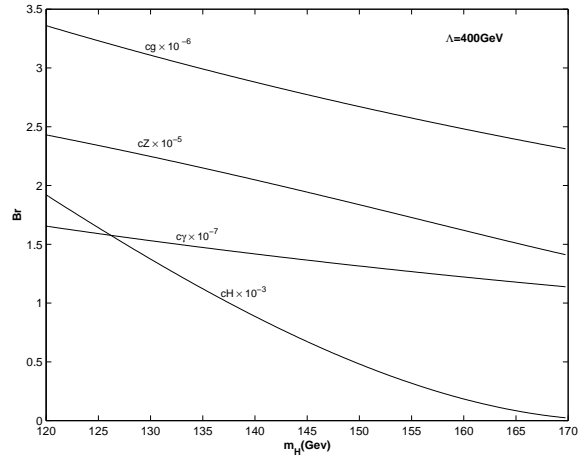}
\caption{\label{Br(mH)}$t\to c V_i$ and $t \to cH$ branching
ratios as functions of $m_H$ for $\Lambda=400$ GeV.}
\end{figure}

\begin{figure}[!htb]
\centering
\includegraphics[width=3in]{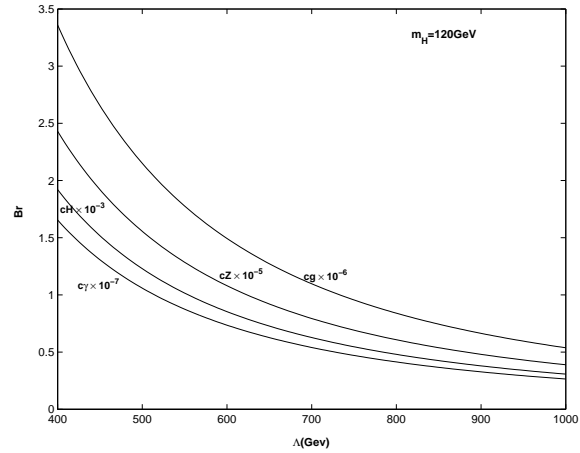}
\caption{\label{Br(Lambda_1)}$t\to c V_i$ and $t \to cH$ branching
ratios as functions of $\Lambda$ for $m_H=120$ GeV.}
\end{figure}

\begin{figure}[!htb]
\centering
\includegraphics[width=3in]{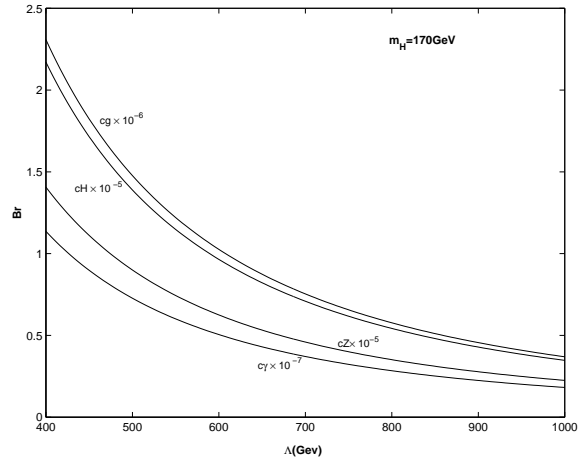}
\caption{\label{Br(Lambda_2)}The same as in Fig.
\ref{Br(Lambda_1)} but for $m_H=170$ GeV.}
\end{figure}

The $t\to c V_i$ and $t\to cH$ branching ratios depend on $m_H$
and $\Lambda$, for which we will consider the ranges $120$ GeV
$\leq m_H\leq 170$ GeV and 400 GeV $\leq \Lambda \leq 1000$ GeV.
In Fig. \ref{Br(mH)}, we show the $t\to c V_i$ and $t\to cH$
branching ratios  versus the Higgs boson mass in the scenario in
which $\Lambda=400$ GeV. Since they are proportional to
$1/\Lambda^2$, the results for other values of $\Lambda$ can be
easily obtained from this figure. We can see that all these
branching ratios, but $Br(t\to cH)$, vary smoothly in the range
considered for the Higgs boson mass. The most pronounced variation
of this channel is due to phase space. We can also observe that
the branching ratios for the $t \to cg$, $t\to c\gamma$, $t\to
cZ$, and $t\to cH$ channels can reach the maximal values
$3.4\times 10^{-6}$, $1.7\times 10^{-7}$, $2.4\times 10^{-5}$, and
$2\times 10^{-3}$, respectively. Finally, Figs. \ref{Br(Lambda_1)}
and \ref{Br(Lambda_2)} show these branching ratios as functions of
$\Lambda$ for $m_H=120$ GeV and $m_H=170$ GeV.

It is worth comparing our results with those obtained in some
specific models. Although in the SM the FCNC top quark decays are
strongly suppressed, they may be considerably enhanced in some of
its extensions \cite{TOPP}. For instance, in the THDM-III the
$t\to cV_i$ and $t\to cH$ (with $H$ a SM-like Higgs boson) decays
can have large branching ratios \cite{REINA2,THDMH}, which are
several orders of magnitude above the SM ones. SUSY models with
universal soft breaking predict branching ratios which are of the
same order of magnitude than the SM ones, but this situation is
improved when the universality is relaxed by allowing a large
flavor mixing between the second and third families, in which case
branching ratios as large as $Br(t\to cg)\sim 10^{-6}$, $Br(t\to
c\gamma)\sim 10^{-8}$, and $Br(t\to cZ)\sim 10^{-8}$ can be
reached \cite{MSSM}, which however are still too small to be
detected. On the contrary, SUSY models with broken $R$-parity may
yield enhanced FCNC top quark decays \cite{SUSYV,SUSYH}. This has
been summarized in Table \ref{Table_Br}, where we show our
predictions for the FCNC top quark decays along with those
obtained in some specific models. Compared with the THDM-III
predictions, the ELA prediction for the $t\to cg$ branching ratio
is almost two orders of magnitude lower, whereas $Br(t\to
c\gamma)$ is of the same order, and $Br(t\to cZ)$ is larger by
more than one order of magnitude. In contrast, the ELA prediction
for $Br(t\to cH)$ is three orders of magnitude larger than in the
THDM-III. As far as SUSY models with broken $R$-parity are
concerned, their predictions for the $t\to cV_i$ decays are all
higher than the ELA results, but the respective prediction for the
$Br(t\to cH)$ is below.

\begin{table}
\caption{\label{Table_Br} Branching ratios for the $t\to cH$ and
$t\to cV_i$ decays in the SM and some of its extensions. The
effective Lagrangian  predictions are displayed in the last two
columns for $\Lambda=400$ and $1000$ GeV. The values shown in each
column correspond to $m_H=120$ GeV and $m_H=170$ GeV,
respectively.}
\begin{center}
\begin{tabular}{|c|c|c|c|c|c|c|c|c|}
\hline Decay & SM & THDM-III & SUSY & ELA ($\Lambda=400$) & ELA ($\Lambda=1000$) \\
\hline \hline $t\to cg$ & $5 \cdot 10^{-11}$ & $10^{-4}-10^{-8}$ &
$\sim 10^{-3}$ & $3.4\times 10^{-6}-2.3\times 10^{-6}$
& $5.4\times 10^{-7}-3.7\times 10^{-7}$  \\
\hline  $t\to c\gamma$ & $5 \times 10^{-13}$ & $10^{-7}-10^{-12}$
& $\sim 10^{-5}$  & $1.7\times 10^{-7}-1.1\times 10^{-7}$
 & $2.6\times 10^{-8}-1.8\times 10^{-8}$ \\
\hline  $t\to cZ$ & $\sim 10^{-13}$ & $10^{-6}-10^{-8}$ & $\sim
10^{-4}$ & $2.4\times 10^{-5}-1.4\times 10^{-5}$
& $3.9\times 10^{-6}-2.3\times 10^{-6}$ \\
\hline  $t\to cH$ & $10^{-14}-10^{-13}$ & $\sim 10^{-6}$ & $\sim
10^{-4}$ &  $2\times 10^{-3}-2.2\times 10^{-5}$
& $3.1\times 10^{-4}-3.5\times 10^{-6}$ \\
\hline
\end{tabular}
\end{center}
\end{table}

It is important to comment on the main differences between the
scenario arising in those models with an extended scalar sector
and that scenario considered so far, in which Higgs-mediated FCNCs
arise from virtual effects of heavy particles. As already
mentioned, the most popular example of those models with extended
scalar sectors is the THDM considered in Ref. \cite{REINA1}. In
that model, dubbed THDM-III, it is assumed that the SM-like Higgs
boson couples diagonally to the fermions at the tree level, so the
$t\to cH$ decay proceeds at the one-loop level due to the exchange
of virtual $h$, $A$ and $H^\pm$ Higgs bosons, which in turn do
have nondiagonal couplings to the fermions. As a consequence, the
respective branching ratio for the $t\to cH$ decay is lower than
in the scenario considered in this work. Although in the THDM-III
the $t\to ch$ and $t\to cA$ decays can arise at tree-level and may
have large branching ratios, it is possible that the $h$ and $A$
Higgs bosons were so heavy that these decays would not be
kinematically allowed. As for the $t\to cV_i$ decays, they are
induced by loops carrying the $h$ and $A$ Higgs bosons. The
Feynman diagrams are similar to those shown in Fig.
\ref{Diagrams}, although there is no contribution from the (iv)
and (v) diagrams, which are not induced in the model considered in
Ref. \cite{REINA1} because the neutral Higgs bosons responsible
for FCNC effects do not couple to the $Z$ boson at tree level as
they do not receive a VEV.

We now would like to discuss on the possible detection of the FCNC
top quark decays at the LHC, which will operate as a top quark
factory, with a production of about $10^{8}$ $t\bar{t}$ events per
year. The dominant mechanism for top quark pair production is
through $q\bar{q}$ or $gg$ annihilation, whereas single top quark
events can be produced through electroweak processes such as $Wg$
fusion or the production of a virtual $W$ boson decaying into
$t\bar{b}$. In this case, the cross section is about $1/3$ that of
$t\bar{t}$ production. Although the observability of a particular
channel decay depends on several factors, in a purely statistical
basis those channels with branching ratios larger than about
$10^{-6}-10^{-7}$ do have the chance of being detected. However,
background problems and systematics may reduce this value by
several orders of magnitude depending on the particular signature.
For instance, the $t\to cg$ mode would require a large branching
ratio in order to be detected as it is swamped by hadronic
backgrounds. As far as the $t\to c\gamma$, $t\to cZ$, and $t\to
cH$ decays are concerned, they could be detected even with
relatively small branching ratios because they would be produced
in a cleaner environment. The LHC will have a sensitivity of about
$2\times 10^{-4}$ and $3.4\times 10^{-5}$ to the $t\to cZ$ and
$t\to c\gamma$ decays, respectively \cite{TOPP}, whereas the
sensitivity to $t\to cH$ can be up to $6.5\times 10^{-5}$
\cite{AS}. From the results presented in Fig. \ref{Br(mH)} and
Table \ref{Table_Br} we can conclude that the $t\to c\gamma$ and
$t\to cZ$ decays would hardly be detected at the LHC. A similar
situation is expected for the $t\to cg$ mode due to background
problems, but the $t\to cH$ decay seems to be more promising. As
far as the FCNC top transitions involving the $u$ quark are
concerned, they are suppressed by a factor of $m_u/m_c$ and are
far from the reach of the LHC.

Finally, we consider that our estimation for the strength of the
$tq'H$ vertex, in which the $\Lambda$ scale was introduced instead
of the Fermi one, is realistic since it describes more
appropriately any possible scenario arising from the underlying
physics responsible for FCNC effects. We believe that this is an
interesting situation as a deep link between flavor physics and
symmetry breaking is possible, thereby favoring this type of
processes.

\section{Conclusions}
\label{c} The copious production of top quark events expected at
the LHC, together with the possibility of detecting the Higgs
boson at this collider, constitute an incentive for studying
Higgs-mediated FCNC top transitions. Due to the large mass of the
top quark and the Higgs boson, the top-Higgs dynamics is expected
to provide a unique scenario to probe the physics beyond the
electroweak scale. This possibility has been explored in a
model-independent manner using the effective Lagrangian technique.
Under the assumption that FCNC top transitions are predominantly
induced by the Higgs boson, the most general Yukawa sector
extended with dimension-six operators, which generates the most
general CP-even and CP-odd $tq'H$ vertex, was studied. We adopted
a slightly modified version of the Cheng-Sher ansatz to estimate
the size of the $tq'H$ vertex. This ansatz comprises three scales:
$m_{q'}$, $m_t$, and the new physics scale $\Lambda$. The most
promising results are obtained when $\Lambda$ is close to the
Fermi scale. The main differences with the scenario arising in
models with extended scalar sectors were emphasized. One of the
most remarkably features of the scenario considered in this work
is the fact that the $t\to q'H$ decay (with $H$ the SM Higgs
boson) arises at the tree level. It turns out that the top quark
decay widths depend only on the Higgs boson mass and the new
physics scale, and for intermediate values of $\Lambda$ they do
not change appreciably when $m_H$ ranges from 120 GeV to 170 GeV.
In such a scenario, the $t\to cg$, $t\to c\gamma$, and $t\to cZ$
decays have branching ratios several orders of magnitude larger
than the ones predicted by the SM, but not large enough to be
detected at the LHC. As far as the $t\to cH$ decay is concerned,
its branching ratio may be up to $10^{-3}$, which is at the reach
of the LHC. This result is three orders of magnitude larger than
the THDM-III prediction and one order larger than in SUSY models
with broken $R$-parity. As for the decays with the $c$ quark
replaced by the $u$ one, the respective branching ratios are
smaller by a factor of $m_u/m_c$, and thus they would  be out of
the LHC reach.

\acknowledgments{We acknowledge support from Conacyt and SNI (M\'
exico). The work of G.T.V is also supported by SEP-PROMEP.

\appendix
\section{$tq'Z$ loop amplitudes}
The scalar functions and the coefficients $f^a_i$ and $g^a_i$
appearing in the $tq'Z$ loop amplitudes are

\begin{eqnarray}
B_0(1)&=& B_0(0,0,m^2_tx^2),\nonumber \\
B_0(2)&=&B_0(m^2_t,0,m^2_ty^2),\nonumber \\
B_0(3)&=& B_0(0,m^2_t,m^2_tx^2),\nonumber \\
B_0(4)&=&B_0(0,m^2_ty^2,m^2_t),\nonumber \\
B_0(5)&=&B_0(m^2_tx^2,m^2_t,m^2_t),\nonumber \\
B_0(6)&=&B_0(m^2_t,m^2_t,m^2_tx^2),\nonumber \\
B_0(7)&=&B_0(m^2_tx^2,m^2_ty^2,m^2_tx^2),\nonumber \\
B_0(8)&=&B_0(m^2_t,m^2_ty^2,m^2_t),
\end{eqnarray}

\begin{eqnarray}
C_0(1)&=&C_0(m^2_t,0,m^2_tx^2,m^2_t,m^2_ty^2,m^2_t),\nonumber \\
C_0(2)&=&C_0(m^2_t,0,m^2_tx^2,m^2_tx^2,m^2_t,m^2_ty^2),\nonumber \\
C_0(3)&=&C_0(m^2_t,0,m^2_tx^2,m^2_ty^2,0,m^2_tx^2),
\end{eqnarray}

\begin{eqnarray}
f^1_0&=&\frac{1}{2}\chi^2,\nonumber\\
f^1_1&=&2\chi^2,\nonumber \\
f^1_2&=&2(1-x^4),\nonumber\\
f^1_7&=&-4x^2\chi ,\nonumber \\
f^1_4&=&\frac{1}{2}\chi \left[x^2(y^2-3)-2(2y^2-3)\right],\nonumber \\
f^1_5&=&-\frac{1}{2}\left[x^4-(2y^2+5)x^2-2(2y^2-5)\right],\nonumber \\
f^1_8&=&-\frac{1}{2}\left[x^4(y^2-4)+x^2(14-3y^2)+8(y^2-2)\right],
\end{eqnarray}

\begin{eqnarray}
g^1_1&=&2x^4+x^2(y^4+2y^2-7)+2(y^2-2)^2,\nonumber \\
g^1_2&=&2\chi^3,\nonumber\\
g^1_3&=&2x\chi\left(x^4+\chi y^2+\right),
\end{eqnarray}

\begin{eqnarray}
f^2_0&=&\frac{3}{2}x^2\chi ^2,\nonumber \\
f^2_1&=&-2x^2\chi ^2,\nonumber \\
f^2_2&=&-2x^2(x^4-1),\nonumber \\
f^2_3&=&\chi ^3,\nonumber \\
f^2_4&=&-\frac{1}{2}\chi \left[-x^4(y^2-3)+2x^2(y^2-4)+2(y^2+1)\right],\nonumber \\
f^2_5&=&-\frac{1}{2}x^2(x^2+2)(x^2-2y^2+1),\nonumber \\
f^2_6&=&-3x^4+x^2(4y^2-3)+2y^2,\nonumber \\
f^2_7&=&4x^6-x^4(y^2-2)-5x^2y^2,\nonumber \\
f^2_8&=&\frac{1}{2}\left[x^6(y^2-4)+x^4(14-3y^2)-2x^2(y^2+4)-2(y^2-2)\right],
\end{eqnarray}

\begin{eqnarray}
g^2_1&=&-2x^6+x^4(y^4-2y^2+7)+2x^2(y^4-2y^2-2)+2,\nonumber \\
g^2_2&=&2\left[-x^8+x^6(y^2+3)-4x^4y^2+x^2(2y^4-3y^2+1)+y^4\right],\nonumber \\
g^2_3&=&-2x^2\chi (x^4+x^2y^2-y^2+1),
\end{eqnarray}

\begin{eqnarray}
f^3_0&=&\frac{1}{2}\chi ^2,\nonumber \\
f^3_1&=&2x^2\chi ^2,\nonumber \\
f^3_2&=&4x^2\chi ,\nonumber \\
f^3_6&=&2x^2\chi ^2,\nonumber \\
f^3_4&=&-\frac{1}{2}\chi \left[x^2(y^2+1)+2(y^2-2)\right],\nonumber \\
f^3_5&=&\frac{1}{2}x^2(5x^2+6y^2-11),\nonumber \\
f^3_7&=&-4x^4\chi ,\nonumber \\
f^3_8&=&\frac{1}{2}\left[x^4(y^2-4)+x^2(6-5y^2)-2(y^2-2)\right],
\end{eqnarray}

\begin{eqnarray}
g^3_1&=&2x^4(2y^2-1)+x^2(3y^4-10y^2+3)+2,\nonumber \\
g^3_2&=&2x^2(y^2-1)\chi ^2,\nonumber\\
g^3_3&=&2x^2\chi \left[x^2(y^2+1)-y^2+1\right],
\end{eqnarray}

\begin{eqnarray}
f^4_0&=&-\frac{3}{2}\chi ^2,\nonumber \\
f^4_1&=&2x^2\chi ^2,\nonumber \\
f^4_2&=&4x^2\chi ,\nonumber \\
f^4_3&=&-\chi ^3, \nonumber \\
f^4_4&=&\frac{1}{2}\chi \left[x^2(y^2+1)-2(2y^2-1)\right],\nonumber \\
f^4_5&=&-\frac{1}{2}\left[x^4+x^2(6y^2-11)+4\right], \nonumber \\
f^4_6&=&-x^6+3x^4-2x^2(y^2-2)-4y^2,\nonumber \\
f^4_7&=&-8x^4+x^2(5y^2+2)+y^2,\nonumber \\
f^4_8&=&-\frac{1}{2}\left[x^4(y^2-4)-x^2(5y^2-14)-2(y^2+2)\right],
\end{eqnarray}

\begin{eqnarray}
g^4_1&=&-(y^2-1)\left[2x^4+x^2(3y^2-7)+2\right], \nonumber \\
g^4_2&=&-2\left[(y^2-1)x^6-2(y^2-2)x^4+y^2(y^2-5)x^2+2y^4\right],\nonumber \\
g^4_3&=&2x^2\chi \left[x^2(y^2+1)-y^2+1\right].
\end{eqnarray}

\end{document}